\documentclass[aps,pra,twocolumn,showpacs,superscriptaddress,groupedaddress]{revtex4}  
\usepackage{amsmath, amsthm, hyperref, endnotes} 

\usepackage{graphicx}  
\usepackage{dcolumn}   
\usepackage{bm}        
\usepackage{amssymb}   
\usepackage{tikz}

\newcommand{\bd}{b^\dagger}

\newcommand{\fr}{\frac{1}{2}}
\newcommand{\tfr}{\tfrac{1}{2}}

\newcommand{\lds}{{\bf L}\cdot {\bm \sigma}}

\newcommand{\ep}{\epsilon}

\newcommand{\eqn}[1]{\begin{equation} #1\end{equation}}
\newcommand{\al}[1]{\begin{align} #1\end{align}}

\newcommand{\At}{\tilde A}
\newcommand{\Bt}{\tilde B}
\newcommand{\Ct}{\tilde C}

\newcommand{\wt}{\widetilde}

\begin{document}

\title{Noncompact dynamical symmetry of  a spin-orbit coupled oscillator}

\date{\today}

\author{S.M.~Haaker$^{1,}$\footnote[1]{Electronic address: s.m.haaker@uva.nl}, F.A.~Bais$^{1,2}$ and K.~Schoutens}
\address{Institute for Theoretical Physics, University of Amsterdam, Science Park 904, 1090 GL Amsterdam, The Netherlands \\ $^2$ Santa Fe Institutute, Santa Fe, NM 87501, USA}

\begin{abstract}
We explain the finite as well as infinite degeneracy in the spectrum of a particular system of spin-$\tfr$ fermions with spin-orbit coupling in three spatial dimensions. Starting from a generalized Runge-Lenz vector, we explicitly construct a complete set of symmetry operators, which  span a noncompact $SO(3,2)$ algebra. The degeneracy of the physical spectrum only involves an infinite, so called {\it singleton} representation. In the branch where orbital and spin angular momentum are aligned the full representation appears, constituting a 3D analogue of Landau levels. Anti-aligning the spin leads to a finite degeneracy due to a truncation of the singleton representation. We conclude the paper by constructing the spectrum generating algebra of the problem.
\end{abstract}

\maketitle

\section{Introduction}
The rich structure revealed by topological phases of matter inspire the investigation of model systems where such phases can be realized. Well-known examples are the integer and fractional quantum Hall (QH) phases,  \cite{klitzing, tsui} which can be characterized by their topology and for which model states have been constructed in a wide variety starting with the work of Laughlin \cite{laughlin}. Recently, a larger class of topological phases have been proposed and discovered, the time-reversal (TR) invariant topological insulators which can be labeled by topological invariants due to their Bloch wave band structure \cite{hasan, qi}. In the two dimensional case the authors of \cite{bernevig} proposed a continuum TR symmetric model, effectively consisting of two integer QH states of opposite chirality. This year, Li and Wu proposed a three dimensional continuum model for TR invariant topological insulators  \cite{li}. They studied a particular model for spin-$\fr$ fermions in the background of a non-Abelian gauge potential, tuned in such a way that a flat dispersion (3D Landau levels) is achieved. The authors argue that this model has helical Dirac surface modes if open boundary conditions are imposed. These modes are protected by time reversal symmetry and are indicative of a non-trivial 3D topological insulator phase. A number of suggestions for experimental realization have been made, such as in strained semiconductors or in ultra-cold atomic matter with synthetic spin-orbit (SO) coupling (see for example \cite{dalibard,anderson}).

In the paper \cite{li} the authors focus on the construction of the eigenfunctions of the model, revealing some remarkable properties such as a form of quaternionic analyticity.  In this paper we present a  complementary algebraic approach to the problem. A particular goal is to understand the degeneracies in the spectrum on the basis of an underlying symmetry algebra.
In general, compact symmetry algebras lead to finite degeneracies, whereas noncompact symmetries give infinite degeneracies.  
In our work we are confronted with a mixture of finite and infinite degeneracies in the spectrum, posing a puzzle on the nature of the underlying symmetry algebra. This problem was studied in \cite{ui} as well, where the authors conclude that an accidental degeneracy does not always imply an underlying symmetry. We show that there is a finite number of operators commuting with the Hamiltonian which form a nonlinear algebra, which we recognize as a `deformed' $SO(3,2)$ symmetry.  These generators include the spin-${1 \over 2} $ generalization of a pair of symmetry vectors that are reminiscent of the Runge-Lenz vector of the Kepler problem. After a simple rescaling of the operators we obtain the linear noncompact Lie algebra $SO(3,2)$. The so-called {\it singleton} representation of this algebra plays a key role in explaining both the infinite and finite degeneracies that feature in the spectrum.

We remark that, quite generally, an insight into the (maximal) symmetry algebra underlying a (one-body or many-body) quantum problem is quite useful. On the one hand, the representation theory provides a catalogue of possible families of quantum states, on the other hand an algebraic structure may contain clues to an underlying geometric picture. An example featuring both these aspects is the $\mathcal{W}_{1+\infty}$ symmetry of specific quantum Hall phases of 2D electrons \cite{cappelli, flohr}. This symmetry reflects the incompressibility of the quantum Hall quantum liquids and it enables an algebraic organization of edge excitations of these same liquids. 

This paper is organized as follows, in section II we will present the Hamiltonian of our interest. We discuss its spectrum and the degeneracies it has. Section III is devoted to the symmetry algebra of the system. We write down all the symmetry operators in a coordinate independent form and show that they form a nonlinear algebra. Moreover, we show that a simple rescaling of the operators results in a linear algebra and we give the representations of this algebra. In section IV we present the operators that allow us to connect states of different energies, the so-called spectrum generating algebra. We end the paper with the conclusions and some ideas for future research in section V.

\section{Spin-orbit coupled harmonic oscillator}
The model proposed by \cite{li} to describe a continuous three dimensional topological insulator is a spin-$\fr$ fermion in a 3D harmonic potential and with a spin-orbit coupling of fixed strength. The Hamiltonian reads 
\begin{equation}
H=\frac{\bm{p}^2}{2m}+\fr m\omega^2\bm{r}^2-\omega\lds,
\label{eq:ham}
\end{equation}
where $L_i$ is the usual orbital angular momentum and $\sigma_i$ are the Pauli matrices.The model is mathematically equivalent to a spin-$\fr$ particle minimally coupled to a static external $SU(2)$ gauge field plus a particular scalar potential,
\eqn{
H=\frac{1}{2m}\left(\bm{p}-q\bm{A}\right)^2+V(r),
}
where the components of the vector potential are $A_i=\fr \omega \ep_{ijk}\sigma_j r_k$ and the harmonic potential is $V(r)=-\fr m\omega^2r^2$. Note that the components of $\bm{A}$ are two by two matrices. Since these components do not commute with each other, this is a so-called non-Abelian gauge field. The field strength associated to this gauge field points in the radial direction and grows with $r$. This gauge potential can be seen as the 3D version of two different 2D ones. For fixed radius $r=1$, this is a spin-$\fr$ particle confined to $S^2$ in a perpendicular magnetic field, resulting in non-Abelian Landau levels on the sphere \cite{estienne}. In $\mathbb{R}^2$ it describes two decoupled layers of quantum Hall states where the two types of particles feel a opposite perpendicular magnetic field, which was mentioned in the introduction \cite{bernevig}. 

\begin{figure}
\includegraphics[scale=0.7]{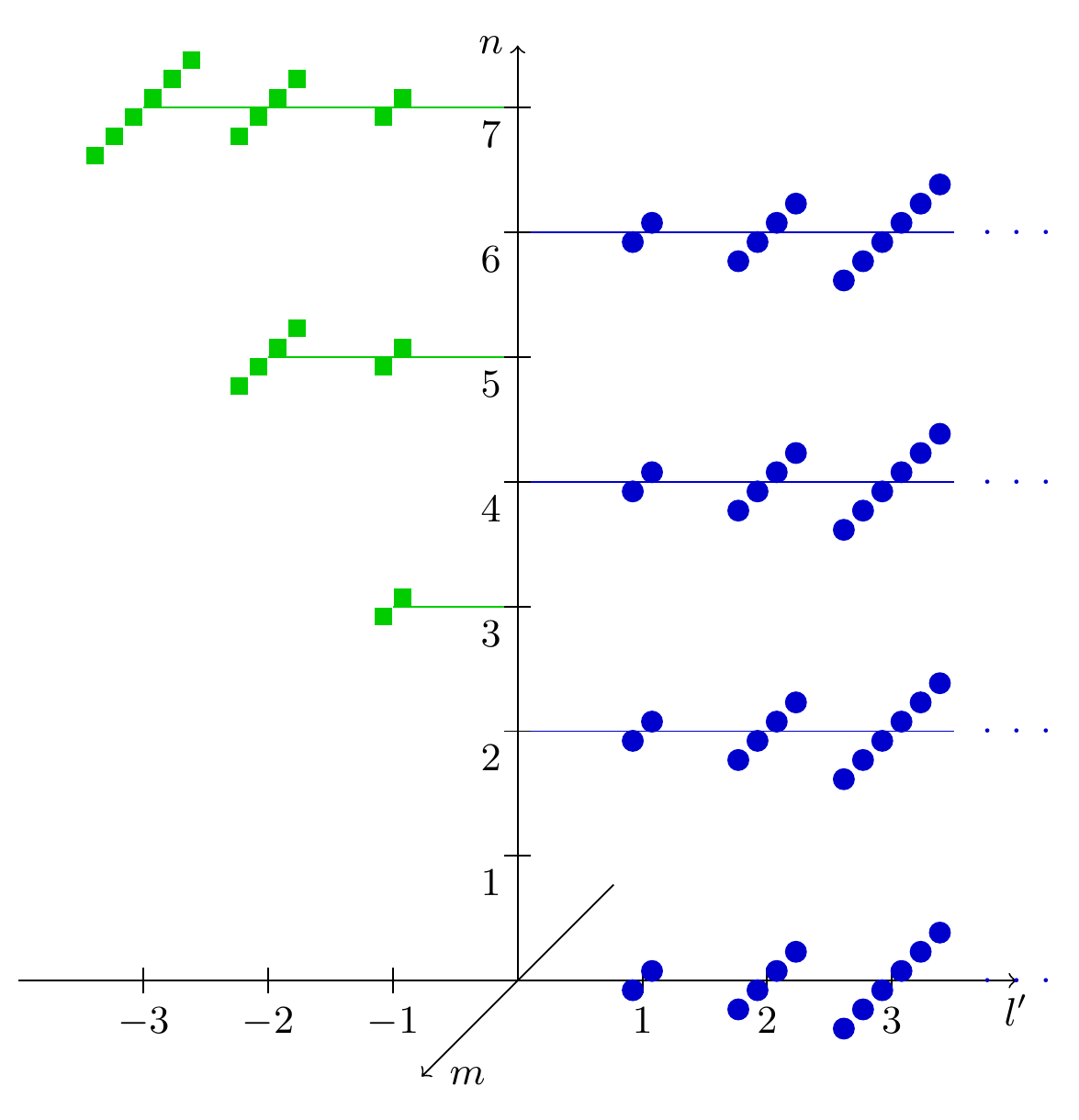}
\caption{(Color online) The spectrum of the spin-orbit coupled harmonic oscillator. The blue dots represent the states of the +branch. At each energy level they are infinitely degenerate with only a lower bound at $l'=1$.  The green squares represent a state of the finitely degenerate --branch. In both cases, for every value of $l'$ there is at most one $SU(2)$ multiplet.}
\label{fig:spectrum}
\end{figure}

In the remainder of this paper we will use the notation in the form of the 3D spin-orbit coupled harmonic oscillator given in Eq.~(\ref{eq:ham}) and we will work in units where $m=1/2$, $\omega=1$, $\hbar=1$. Since this is a single-particle radial problem there are several ways of solving the system. As we are interested in the algebraic approach we will look for operators that commute with the Hamiltonian. First of all, $H$ commutes  with the total angular momentum operators ${\bf J}={\bf L}+\fr \bm{\sigma}$. So the Hilbert space will arrange into $SU(2)$ multiplets, where every irreducible representation may be labeled by its $\bm{J}^2$ eigenvalue $j_\pm(j_\pm+1)$. Here $j_\pm=l\pm \fr$ and $l=0,1,...$ is associated with the orbital angular momentum $\bm{L}^2$. We can diagonalize in $\bm{J}^2$ and $J_3$ as is standard in the $SU(2)$ case, but we will choose a slightly different convention.  Instead, we first define $A_3\equiv \lds+1$. This operator commutes with $H$ and with $J^2$ and its eigenvalue can be easily obtained from the relation $A_3={\bf J}^2-{\bf L}^2+\frac{1}{4}$. We will label the eigenstates of $H$ by their $A_3$ and $J_3$ eigenvalues
\begin{equation}
A_3\,\psi_{n,l',m}=l'\,\psi_{n,l',m} \ , \qquad   J_3\,\psi_{n,l',m}=m\,\psi_{n,l',m} \ , \label{cartansub}
\end{equation}
where $l'=\pm1, \pm 2,...$ and $-(|l'|-\fr)\leq m\leq |l'|-\fr$. The eigenvalues of $J^2$ in terms of $l'$ follow from the relation $J^2=A_3^2-\frac{1}{4}$. In section \ref{sectionSGA} we will derive the spectrum by constructing energy ladder operators. At this point we will solve the Schr\"odinger equation, giving us the spectrum and the energy eigenstates. If we switch to spherical coordinates and use separation of variables the angular part is a linear combination of spherical harmonics $Y_{lm}(\hat{\Omega})$
\begin{align}
\chi^+_{lm}&=\sqrt{\frac{l+m+1}{2l+1}}Y_{lm}\left(\begin{matrix} 1\\0\end{matrix}\right)+\sqrt{\frac{l-m}{2l+1}}Y_{l,m+1}\begin{pmatrix}0\\1\end{pmatrix} \nonumber \\
\chi^-_{lm}&=\sqrt{\frac{l-m}{2l+1}}Y_{lm}\begin{pmatrix} 1\\0\end{pmatrix}-\sqrt{\frac{l+m+1}{2l+1}}Y_{l,m+1}\begin{pmatrix}0\\1\end{pmatrix},
\end{align}
where the spin states are diagonal in $\sigma_z$. Note, that we momentarily switched to labeling the states by $l$ and $\pm$, where ${\bf L}^2\chi_{lm}^\pm=l(l+1)\chi_{lm}^\pm$. The (unnormalized) radial part of the eigenstates is the same for both $\pm$ states 
\begin{equation}
R_{kl}(r)=r^le^{-r^2/4}L(-2k, 2l+\frac{5}{2}, r^2/2).
\end{equation}
$L$ is a generalized Laguerre polynomial, which in this particular case has a finite number of terms.
The spectrum is the following
\begin{equation}
E=\left\{\begin{array}{ll}
2k+\frac{3}{2} & \quad l'>0\\
2k-2l'+\frac{5}{2} & \quad l'<0\end{array}\right.,
\end{equation}
where $k=0,1,\ldots$. The energy does not depend on $m$, as it should not, because of conservation of total angular momentum ${\bf J}$. But there is a bigger (accidental) degeneracy in the system. For the +branch, the energy also does not depend on $l'$, resulting in an infinite degeneracy at every energy level. On the other hand, the --branch has a degeneracy in $l'$, but it is finite. We can express the energy in terms of a new quantum number $n$ as
\begin{equation}
E=n+\tfrac{3}{2}, \qquad \left\{ \begin{array}{ll}
l'>0: &n=0,2,\ldots\\
l'<0: &n=3,5,\ldots\end{array}\right..
\end{equation}
For $n$ even, the energy levels have an infinite degeneracy and for odd $n$, the allowed values of $l'$ are $l'=-1,\ldots, \frac{1}{2}(1-n)$. The spectrum is depicted in Fig.~\ref{fig:spectrum}. We want to stress that the infinite degeneracy of the +branch is only present when the strength of the spin-orbit coupling term in the Hamiltonian is exactly $-\omega$. (If the strength is exactly opposite, i.e. $+\omega$, there would be an infinite degeneracy in the --branch.)

\section{Symmetry algebra}
The aim of this paper is to understand the degeneracies in the spectrum from an underlying symmetry algebra. The authors of \cite{ui} raised the question if an accidental degeneracy always implies a symmetry algebra. They investigated the degeneracies of the Hamiltonian in Eq. \eqref{eq:ham} and concluded  that there was no such algebra. Their conclusions were mainly based on the fact that a Lie algebra has either only finite dimensional nontrivial irreducible representations or only infinite dimensional ones, depending on if it is a compact or a noncompact algebra, respectively. Clearly, the spectrum of $H$ contains both finite and infinite dimensional representations, which is indeed a puzzle. Moreover, they constructed operators that connect different $SU(2)$ multiplets within one energy level by mapping $SU(2)$ highest weight states onto each other, but these operators do not commute with $H$.

We will show that it is possible to construct operators that commute with $H$ and couple the different $SU(2)$ multiplets. And we will show that these operators have an underlying noncompact Lie algebra structure. Consider the following two vectors of Hermitian operators
\begin{align}
\widetilde M_i&=\tfrac{1}{4}\left(r_iA_3+A_3 r_i\right)+\tfr\left(({\bf p}\times {\bf J})_i-({\bf J}\times {\bf p})_i\right) \nonumber \\
\widetilde N_i&=\tfrac{1}{2}\left(p_iA_3+A_3 p_i\right)-\tfrac{1}{4}\left(({\bf r}\times {\bf J})_i-({\bf J}\times {\bf r})_i\right) \ .
\end{align} 
Note the symmetry in the two definitions above under the  map $r_i \rightarrow  2p_i$ and $p_i \rightarrow -\tfrac{1}{2}r_i $, which is in fact of order four.  We remark that the Hamiltonian has not just this discrete  symmetry but the continuous $U(1)$ version of it. One may explicitly show that $\widetilde {\bf M}$ and $\widetilde {\bf N}$ both transform as vectors under $\bm{J}$ and that they commute with the Hamiltonian. Moreover they connect different $SU(2)$ irreducible representations with each other, so they are exactly the operators that we were looking for. These operators are spin-$\fr$ generalizations the well-known Runge-Lenz vector present in the Kepler problem. 

We wish to determine the algebra formed by $\widetilde {\bf M}$ and $\widetilde {\bf N}$ together with $\bm{J}$ and $A_3$. When we compute the commutation relations of these operators amongst themselves we run into terms nonlinear in $H$ and $A_3$. The explicit commutation relations are given in the Appendix. The nonlinearity encountered here is very similar to what happens in the Kepler problem. In that case the Runge-Lenz vector $\bm{A}^{RL}$ transforms as a vector under orbital angular momentum $\bm{L}$, but the commutation relations with itself are $[A^{RL}_i, A^{RL}_j]=-i\ep_{ijk}2H^KA^{RL}_k$, where $H^K$ is the Hamiltonian of the Kepler problem. The vector needs to be rescaled by $(-2H^K)^{-\fr}$ in order to obtain the $SO(4)$ commutation relations.

With this in mind we set out to find an appropriate rescaling factor, which would enable us to get more grip on the problem. The easiest way to find the appropriate rescaling is by looking at the action of the symmetry operators on an energy eigenstate $\psi_{n,l',m}$. In order to do so we write the operators on a Cartan basis. First redefine 
\begin{align}
\widetilde M_\pm&=\tfrac{1}{\sqrt{2}}(\widetilde M_x\pm i \widetilde M_y) \nonumber \\
 \widetilde N_\pm&=\tfrac{1}{\sqrt{2}}(\widetilde N_x\pm i \widetilde N_y).
\end{align}
As Cartan subalgebra we obviously choose $\{ J_3, A_3\}$ and the operators corresponding to the root vectors are $J^\pm$ and 
\begin{align}
\wt A_+&= \tfrac{1}{\sqrt{2}}(\widetilde M_z- i \widetilde N_z) \nonumber  \\
\wt A_-&=\tfrac{1}{\sqrt{2}}(\widetilde M_z+ i\widetilde N_z) \nonumber  \\
\wt B_+&=-\tfrac{1}{\sqrt{2}} (\widetilde M_+- i\widetilde N_+)\nonumber  \\
\wt B_-&=-\tfrac{1}{\sqrt{2}}(\widetilde M_-+ i\widetilde N_-) \nonumber  \\
\wt C_+&=\tfrac{1}{\sqrt{2}} (\widetilde M_-- i\widetilde N_-) \nonumber  \\
\wt C_-&=\tfrac{1}{\sqrt{2}}(\widetilde M_++ i \widetilde N_+).  \label{Aplus_Cmin}
\end{align}
The action of these operators on an energy eigenstate is
\begin{widetext}
\begin{align}
J_\pm\psi_{n,l',m}&=\sqrt{\tfr(l'-m\mp\tfr)(l'+ m\pm \tfr)}\psi_{n,l',m\pm1}  \nonumber \\
\wt A_\pm\psi_{n,l',m}&=\sqrt{\tfr(l'-m\pm\tfr)(l'+ m\pm\tfr)(n+2l'\pm1)}\psi_{n,l'\pm1,m} \nonumber  \\
\wt B_\pm\psi_{n,l',m}&=\frac{l'}{2|l'|}\sqrt{(l'+m\pm\tfr)(l'+ m\pm\tfrac{3}{2})(n+2l'\pm1)}\psi_{n,l'\pm1,m\pm1} \nonumber \\
\wt C_\pm\psi_{n,l',m}&=\frac{l'}{2|l'|}\sqrt{(l'-m\pm\tfr)(l'- m\pm\tfrac{3}{2})(n+2l'\pm1)}\psi_{n,l'\pm1,m\mp1}. \label{cpm}
\end{align}\end{widetext}

Now it is quite easy to guess the right rescaling operator. Consider the operator $F=H+2A_3-\frac{5}{2}$, which commutes with $H, A_3$ and $J_i$ and has the following eigenvalues
\begin{align}
F\psi_{n,l',m}&=(n+2l'-1)\psi_{n,l',m}\label{F-} \ .
\end{align}
The operators in (\ref{Aplus_Cmin}) commute with $F$ as $[F, \wt A_\pm]=\pm 2\wt A_\pm$ , and similarly for $\wt B_\pm$ and $\wt C_\pm$. Now rescale them as $A_+=\frac{1}{\sqrt{F}}\wt A_+$ and $A_-=\wt A_- \frac{1}{\sqrt{F}}$ and again the same way for $\wt B_\pm$ and $\wt C_\pm$. Note that the order of the operators is important to ensure that $A_+^\dagger=A_-$ and that the right factor is obtained when acting on an energy eigenstate. For the scaling factor $F$ to be well defined we need to make sure that \makebox{$F>0$} for all states in the Hilbert space. This condition is met for the states of the +branch, but the \mbox{--branch} includes states for which $F=0$.  We will address this point after discussing the representations of $SO(3,2)$. 

\begin{figure}[t]
\includegraphics[scale=1.1]{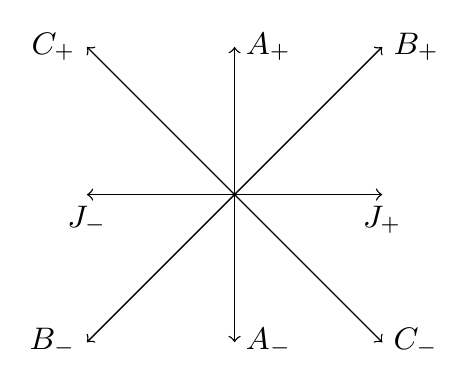}
\caption{Root diagram of the rescaled algebra corresponding to $SO(3,2)$. It is of rank 2, where we choose as Cartan subalgebra $\{J_3, A_3\}$.}
\label{rootdiagram}
\end{figure}

 From the definitions of $J_i, A_3, A_\pm, B_\pm$ and $C_\pm$ we can explicitly compute the commutation relations of the rescaled symmetry operators, by acting on an energy eigenstate. They form a 10-dimensional (linear) Lie algebra of rank 2, corresponding to the noncompact algebra $SO(3,2)$. The corresponding group consists of transformations that leave the quadratic form $x_1^2+x_2^2+x_3^2-x_4^2-x_5^2$ invariant. As said before, we choose as Cartan subalgebra  $\{J_3, A_3\}$, which both are compact generators, leading to the root diagram shown in Fig.~\ref{rootdiagram}. The nonzero commutators are
\begin{align}
[A_3, A_\pm]&=\pm A_\pm \quad & [J_3,J_\pm]&=\pm J_\pm \nonumber \\
[A_3,B_\pm]&=\pm B_\pm \quad &[J_3,B_\pm]&=\pm B_\pm \nonumber  \\
[A_3,C_\pm]&=\pm C_\pm \quad &[J_3,C_\pm]&=\mp C_\pm \nonumber  \\
[J_\pm, A_\pm]&=\pm B_\pm \quad& [J_\mp,B_\pm]&=\pm A_\pm \nonumber \\
[J_\mp, A_\pm]&=\pm C_\pm \quad &[J_\pm,C_\pm]&=\pm A_\pm \nonumber \\
[A_\mp, B_\pm]&=\pm J_\pm \quad &[A_\pm, C_\mp]&=\mp J_\pm \nonumber \\
[J_+,J_-]&=J_3\quad &[B_+,B_-]&= -(A_3+J_3) \nonumber \\
[A_+,A_-]&=-A_3\quad& [C_+,C_-]&= -(A_3-J_3) .
\end{align}
Since the symmetry operators $\{J_i, A_3, \widetilde M_i, \widetilde N_i\}$ and the rescaling operator $F$ are Hermitian, we need to study the unitary representations of $SO(3,2)$, if we want to describe the spectrum in Fig. \ref{fig:spectrum}. As we mentioned before, it is well known that the nontrivial unitary irreducible representations of a noncompact group are all infinite dimensional. The unitary infinite dimensional representations of the covering group of $SO(3,2)$ have been studied by a number of authors \cite{ehrman, dirac}, while they were completely classified by Evans \cite{evans}. One of the reasons why they attracted a lot of attention is because $SO(3,2)$ and $SO(4,1)$ are the only simple Lie groups which can be contracted to the Poincar\'e group. 

Following Evan's notation, a representation can be labeled by its {\it extremal weight} $(q, s)$ in the same way that one usually labels a representation of a compact group by its highest weight. The lower bound of $A_3$ is indicated by $q$ and the lowest $SU(2)$ multiplet of the irreducible representation is labeled by $s$. The representation that forms the +branch have $q=1$ and $s=\fr$.
This representation was first found by Ehrman \cite{ehrman} and later Dirac wrote an explicit form of this representation \cite{dirac}, which is known as the {\it Dirac singleton} of spin-$\tfr$. It is one of the singleton representations, because all its weights have unit multiplicity. Whereas representations of compact groups are uniquely labeled by the value of the Casimir operators, noncompact groups might allow more than one inequivalent representation. This is the case for the representation that we are considering.

For completeness we give the action of the $SO(3,2)$ operators on the  eigenstates of $H$, which are indeed in exact agreement with those on the states of the singleton representation.
\begin{align}
J_\pm\psi_{n,l',m}&=\sqrt{\tfr(l'-m\mp\tfr)(l'+ m\pm \tfr)}\psi_{n,l',m\pm1} \nonumber \\
A_\pm\psi_{n,l',m}&=\sqrt{\tfr(l'-m\pm\tfr)(l'+ m\pm\tfr)}\psi_{n,l'\pm1,m} \nonumber \\
B_\pm\psi_{n,l',m}&=\frac{l'}{2|l'|}\sqrt{(l'+m\pm\tfr)(l'+ m\pm\tfrac{3}{2})}\psi_{n,l'\pm1,m\pm1} \nonumber \\
C_\pm\psi_{n,l',m}&=\frac{l'}{2|l'|}\sqrt{(l'-m\pm\tfr)(l'- m\pm\tfrac{3}{2})}\psi_{n,l'\pm1,m\mp1}.\label{cpm}
\end{align}
The structure and weight multiplicities of the representations can be well described by studying the decompositions under the various maximal subalgebras of  $SO(3,2)$. The following are the regular subalgebras that share the same Cartan subalgebra:\\
 (a) $SO(2) \times SO(3)$ generated by $A_3$ and $J_i$,\\
 (b) $SO(2,1) \times SO(2)$ generated by $A_3, A_\pm$  and $J_3$,\\
 (c)  $SO(2,1)\times SO(2,1)= SO(2,2)$ generated by $B_\pm, C_\pm$ and $A_3\pm J_3$. \\

\begin{figure*}
\includegraphics[scale=0.75, angle=0]{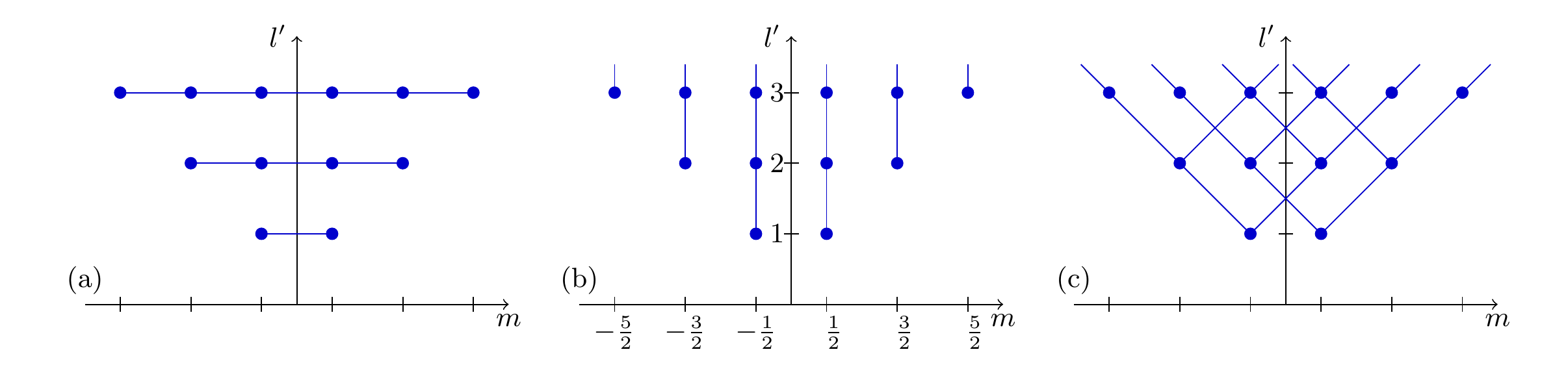}
\caption{(Color online) The three decompositions of the singleton under the subalgebras (a-c). The drawn lines connect the weights corresponding to the irreducible representations of these subalgebras. The figure shows the representation with $l'>0$. The inequivalent representation with $l'<0$ can be obtained by reflecting in the $m$-axis.}
\label{fig:decomp}
\end{figure*}

The weight decomposition under these subalgebras  is shown in Fig.~\ref{fig:decomp} for the representation with $l'>0$. The weight diagram of the inequivalent representation with $l'<0$ is obtained by reflection in the $m$-axis.

It turns out that there is one more regular maximal subalgebra of interest in this problem. Clearly the $SO(3,2)$ has not just the $SO(2,2)$ we just discussed, but also a $SO(3,1)$ subalgebra which is maximal. It is generated by the $J_i$  operators where we add the  rescaled $\widetilde M_i$ operators, denoted by $M_i$ (picking $\widetilde N_i$ instead of $\widetilde M_i$ gives an equivalent representation). The $M_i$ operators can be expressed in terms of the $SO(3,2)$ roots as follows
\begin{equation}
M_z=\tfrac{1}{\sqrt{2}}(A_++A_-), \qquad  M_\pm=\tfrac{1}{\sqrt{2}}(C_\mp-B_\pm) \ .
\end{equation}
Note that we cannot choose the same Cartan subalgebra since $A_3$ is not part of this subalgebra. When we act on the eigenstates of the Hamiltonian, we obtain exactly the same coefficients and multiplicities as described by Harish-Chandra in \cite{harish}, where he constructs the unitary infinite dimensional representations of $SO(3,1)$ in terms of the irreducible representations of the $SO(3)$ subalgebra. From this we may conclude that the degeneracies in the spectrum also form an irreducible representation (principal series) under the $(3+1)$-dimensional Lorentz algebra. Bearing the inclusions $SO(3,2) \supset SO(3,1) \supset SO(3)$ in mind it is not so surprising that the singleton remains irreducible under $SO(3,1)$, in contrast with the situation for $SO(2,2)$ depicted in 
Fig.~\ref{fig:decomp}(iii). The largest symmetry in our problem however remains the $SO(3,2)$ algebra.

Now let us return to the physical unrescaled operators 
\begin{align}
\At_+&=\sqrt{F}A_+=\frac{1}{\sqrt{2}}(\tilde M_z-i\tilde N_z) \nonumber \\
\At_-&=A_-\sqrt{F}=\frac{1}{\sqrt{2}}(\tilde M_z-i\tilde N_z),\label{factorA+-}
\end{align}
where $\Bt_\pm, \Ct_\pm$ are defined in the same fashion. We want to check what the influence is of the rescaling factor $F$ on the singleton representation. In other words, knowing the $SO(3,2)$ representation, how do we obtain the spectrum of the physical system?
First note that the $SU(2)$ multiplets are not affected by the rescaling factor since $\bm{J}$ has not been rescaled and commutes with $F$. So we need to figure out which $SU(2)$ representations are present in the physical spectrum. We mentioned before that $F>0$ for all the states of the +branch, meaning that the coefficients of the physical unrescaled operators will be altered by this factor, but we will still be left with an infinite dimensional representation characterized by the same lower bound $l'=1$ and with the same multiplicities. The \mbox{--branch}, the states with negative $A_3$ eigenvalue, also form an infinite representation under $SO(3,2)$ in this case with an upper bound, $l'=-1$, but the physical lowering operators annihilate the states for $l'=\fr(1-n)$. This creates a lower bound depending on the energy level, resulting in a finite dimensional truncation of the representation. This all perfectly agrees with the spectrum in Fig.~\ref{fig:spectrum}.

We remark that the operator-dependent scale transformations Eq.~(\ref{factorA+-}) are very similar to the so-called Holstein-Primakoff transformations \cite{holstein}. The latter relate bosonic oscillators $\{a,a^\dagger\}$ to $SU(2)$ spin-operators $\{S_\pm, S_z\}$ through
\begin{eqnarray}
& S_+=\sqrt{(2s-a^\dagger a)} \, a, \quad S_-= a^\dagger \sqrt{(2s-a^\dagger a)}, \nonumber \\[2mm]
& S_z=(s-a^\dagger a).
\label{HP}
\end{eqnarray}
The vanishing of $S_-$ acting on the state with $S_z=-s$ arises from the vanishing of the scale factor on this state, which has $a^\dagger a=2s$. We recall that a transformation very similar to Eq.~(\ref{HP}) relates the symmetry operators for 2D Landau levels on the plane to those pertaining to (finite dimensional) Landau levels in a spherical geometry \cite{haldane}. 

\section{Spectrum generating algebra}\label{sectionSGA}
After determining the symmetry algebra and explaining why there are both finite and infinite degeneracies present in this system, we will now describe operators that connect different energy levels, the so-called spectrum generating algebra (SGA) \cite{liwu}. Two operators that commute with $J_3$ and $A_3$ are
\al{
K_+&=-\tfr\bd_i\bd_i \nonumber  \\
K_-&=-\tfr b_i b_i ,
}
where the $\bd_i$ and $b_i$ are the usual bosonic raising and lowering operators
\begin{align}
\bd_i&=r_i/2-ip_i \nonumber \\
b_i&=r_i/2+ip_i \nonumber  \\[2mm]
[b_i, \bd_j]=\delta_{ij}, &\quad [b_i, b_j]=[\bd_i, \bd_j]=0
\end{align}
in terms of which the harmonic oscillator (HO) part of the Hamiltonian can be expressed as $H_{HO}=\bd_i b_i+\frac{3}{2}$.
The operators $K_\pm$ raise or lower the energy by steps of $2$ as can be seen from their commutator
\begin{align}
[H,K_\pm]&=\pm 2K_\pm \nonumber \\
[K_+,K_-]&=-(H+A_3-1) \ .
\end{align}
It is evident that they form an $SO(2,1)$ algebra when we define $K_3=H+A_3-1$. The representation theory of this algebra is well-known \cite{bargmann} and again the unitary irreducible representations are all infinite dimensional, reflecting the fact that the energy is not bounded from above. 
The coefficients of these operators when they act on an energy eigenstate are
\begin{align}
K_\pm\psi_{n,l',m}&=\tfr\sqrt{(n+1\pm1)(n+2l'\pm1)}\;\psi_{n\pm2,l',m} \ .
\end{align}
Unitary irreducible representations of $SO(2,1)$ have either a lower or an upper bound and can be uniquely defined by this bound. For every value of $l'$ and $m$ there is an infinite tower of states, corresponding to one such representation. From the coefficients of $K_\pm$, we see that for the +branch all representations have a lower bound $n=0$ and the representations of the --branch have a lower bound $n=-2l'+1$, in perfect agreement with the physical spectrum.
Note that the algebra $SO(3)\times SO(2,1)$ spanned by $J_i$ and $K_i$ is associated with the radial symmetry of the system, allowing us to write the wave function as a product of a radial and an angular function, see for example \cite{wybourne}. We would like to mention that when we rescale $K_\pm$ by a factor of $\sqrt{(H-\fr)/F}$, these operators commute with the symmetry algebra, resulting in a dynamical algebra $SO(2,1)\times SO(3,2)$. 

We can also construct operators that move between the two branches. Consider $T_+=\sum_ib_i^\dagger\sigma_i$ and $T_-=\sum_ib_i\sigma_i$, in terms of which the Hamiltonian can be expressed as $H=T_+T_-+\frac{3}{2}$ \cite{susy}. These operators anticommute with $A_3$ and map +branch states onto \mbox{--branch} states and vice versa
\begin{align}
T_+\psi_{n,l',m}&=\sqrt{n+2l'+1}\;\psi_{n+2l'+1, -l', m} \nonumber \\
T_-\psi_{n,l',m}&=\sqrt{n}\;\psi_{n+2l'-1, -l', m}.
\end{align}
We have not yet succeeded in extending the $SO(3,2)$ symmetry algebra to a complete spectrum generating (super) algebra including both the $K_\pm$ and $T_\pm$ operators.

\section{Conclusions}
We have explicitly constructed the symmetry algebra of a spin-orbit coupled harmonic oscillator given in \eqref{eq:ham}. Besides the $SU(2)$ symmetry coming from conservation of total angular momentum we have identified six other operators that commute with $H$ and are a spin-generalized version of the Runge-Lenz vector. Commuting these symmetry operators results into non-linear commutation relations, which had to be expected since there are finite and infinite degeneracies in this model. We show that a simple rescaling of the operators leads to linear commutation relations which we recognize as a $SO(3,2)$ algebra. The infinite degenerate branches of the spectrum are the singleton representation of $SO(3,2)$ and the finite degenerate levels are a truncated version of the singleton. We also identify four operators that connect different energy levels with each other, forming the spectrum generating algebra.

Let us conclude the paper with some further comments. Our results suggest a geometrical picture for 3D Landau levels based on 4-dimensional Anti de Sitter space (AdS4), whose isometries are precisely $SO(3,2)$. We have established that, as far as their quantum orbitals are concerned, particles in the 3D Landau levels experience a geometry that is a radial deformation of AdS4 rather than flat space. We expect that this qualitative observation can be made more precise  - for example in the form of accurate statements about magnetic translations.

Another observation is that choosing a plus sign in front of the $\lds$ term in the Hamiltonian does not change much. The degeneracy of the \mbox{--branch} will be infinite and the +branch will be finitely degenerate. These two Hamiltonians are supersymmetric partners. One may in $D=3$ choose a different spin representation in the Hamiltonian, $H=H_0-\alpha {\bf L}\cdot{\bf S}$, where $\alpha$ is a constant. For spin-$s$ particles this gives us $2s+1$ branches, of which we can always make one flat by adjusting the coefficient $\alpha$. Presumably that will not change the fundamental structure, but  will change the representation content to a corresponding higher spin representation of the group $SO(3,2)$. 

A final question of interest is to analyze this system in higher dimensions to determine whether  a generic hierarchy of symmetries  arises similar to what is the case for the ordinary harmonic oscillator.
\begin{acknowledgements}
We thank Jan de Boer and Benoit Estienne for many helpful discussions. KS acknowledges hospitality at Harvard's Physics Department, where part of this work was done. SH is supported by the Stichting voor Fundamenteel Onderzoek der Materie (FOM) of the Netherlands.
\end{acknowledgements}

\appendix*
\section{Commutation relations}\label{app:A}

As mentioned in the body of the text the commutation relations of the unrescaled symmetry operators contain nonlinear terms. We will explicitly give the nonzero commutation relations here. The following five commutation relations are linear and are the same as those of $SO(3,2)$.
\begin{align}
[J_i, J_j]&=i\ep_{ijk}J_k \nonumber \\
[J_i, \wt M_j] &=i\ep_{ijk}\wt M_k \nonumber \\
[J_i, \wt N_j] &=i\ep_{ijk}\wt N_k \nonumber \\
[A_3, \wt M_i]&=-i\wt N_i \nonumber  \\
[A_3, \wt N_i]&=i\wt M_i.
\end{align}
The next three nonzero commutation relations are nonlinear and therefore are not as those of $SO(3,2)$.
\begin{align}
[\widetilde M_i, \wt M_j]&=-i\ep_{ijk}J_k(H+3A_3-\tfrac{3}{2}) \nonumber \\
[\wt N_i, \wt N_j]&= -i\ep_{ijk}J_k(H+3A_3-\tfrac{3}{2}) \nonumber \\
[\wt M_i, \wt N_j]&=i\delta_{ij}A_0(H+3A_3-\tfrac{3}{2}) \nonumber \\
& \quad +\tfrac{1}{4}i\delta_{ij}-\tfrac{i}{2}(J_iJ_j+J_jJ_i).
\end{align}


\vfill

\begin{thebibliography}{99}
\bibitem{liwu}
The authors of \cite{li} also construct operators that connect different energy levels, but they use the radial symmetry of the system and only consider 1D radial operators. Our construction is coordinate independent.
\end{thebibliography}
\end{document}